\documentclass[twocolumn,aps,prl,floatfix]{revtex4-2}
\usepackage{amsmath}
\usepackage{amsfonts}
\usepackage{amssymb}
\usepackage{graphicx}
\usepackage{color}
\usepackage{bm}

\usepackage{soul}

\usepackage{mathtools}

\DeclarePairedDelimiterX\braket[2]{\langle}{\rangle}{#1 \delimsize\vert #2}
\usepackage[final]{hyperref} % adds hyper links inside the generated pdf file
\hypersetup{
	colorlinks=true,       % false: boxed links; true: colored links
	linkcolor=blue,        % color of internal links
	citecolor=blue,        % color of links to bibliography
	filecolor=magenta,     % color of file links
	urlcolor=blue         
}

\newcommand{\new}[1]{\textcolor{blue}{#1}}

\begin{document}

\title{Pontus-Mpemba effects}

\author{Andrea Nava}
\author{Reinhold Egger}
\affiliation{Institut f\"ur Theoretische Physik, Heinrich-Heine-Universit\"at, D-40225  D\"usseldorf, Germany}

\begin{abstract}
Mpemba effects occur after a sudden quench of control parameters if for ``far'' (or ``hot'') initial states with respect to a final target state, 
the relaxation time toward the target state is shorter than for ``close'' (or ``cold'') initial states. Following a strategy of fishermen in Pontus described by Aristotle, we introduce the Pontus-Mpemba effect as a two-step protocol which includes the time needed for preparing the system in the ``far'' initial state
that can now be an arbitrary nonequilibrium state.  Our protocol needs no parameter distance concept and applies to general (classical or quantum) systems. We find that all possible Pontus-Mpemba effects fall into three classes and illustrate the theory for open Markovian two-state quantum systems.
 \end{abstract}
\maketitle

\emph{Introduction.---}Classical \cite{Mpemba_1969,Lu2017,Lasanta2017,Klich2019,Jesi2019,Torrente2019,Santos2020,Megias2022,Chetrite2021,Schwarzendahl2022,Walker2023,Ibanez2024,Adalid2024,Vu2025,Bisson2025} and quantum 
\cite{Nava2019,Carollo_2021,Kochsiek2022,Rylands2023,Nava2024,Murciano_2024,Liu2024,Moroder2024,Ares2025b,Zatsarynna2025,Su2025,Wang2024,Giulio2025,Strachan2024,Turkeshi2024,Qian2025,Chang2025,Lacerda2025,Westhoff2025} versions of the celebrated Mpemba effect (ME) are presently attracting a lot of attention.  This interest is mainly driven by the intriguing physics behind the ME, by recent trapped ion experiments \cite{Joshi2024,Shapira2024,Zhang2025}, and by the 
promise of speeding up state preparation and manipulation protocols.  For recent reviews, see Refs.~\cite{Ares2025,Teza2025}.
According to the standard definition of the ME employed in the works cited above, one considers a single-step protocol where control parameters are suddenly quenched from initial values (defining the
initial system state) to final values (defining the target state). 
Comparing two hand-picked initial parameter sets, the ME occurs if the relaxation times  toward the target state are ordered in a
``counterintuitive'' way.  For example, 
if one considers the relaxation times $\tau_{c,h}$ for systems starting at 
the respective temperatures $T_c<T_h$ and relaxing toward a target state with temperature $T_f<T_c$, the ME happens for $\tau_h<\tau_c$.
Importantly, the ME is a true nonequilibrium phenomenon  absent under slow parameter variations.

However, a different and arguably more stringent definition requires
that starting from a common initial configuration, the \emph{entire two-step warming-and-freezing process} (say, $T_c\to T_h\to T_f$) must be faster than the single-step process of freezing ($T_c\to T_f$).  We call the ME in such a two-step protocol ``Pontus-Mpemba effect'' (PME) since this principle was already noted by Aristotle in his description of fishing strategies in the ancient towns of Pontus: ``Hence many people, when they want to cool hot water quickly, begin by putting it in the sun. So the inhabitants of Pontus when they encamp on the ice to fish (they cut a hole in the ice and then fish), pour warm water round their reeds that it may freeze the quicker, for they use the ice like lead to fix the reeds \cite{Aristotle}.''  Indeed, if the warm-up step $T_c\to T_h$ requires more time than is saved during the subsequent freezing process $T_h\to T_f$, the Pontus fishermen could take no advantage by doing so as compared to the direct freezing route $T_c\to T_f$.  For related work, see Ref.~\cite{Gal2020}.

\begin{figure}
\includegraphics[width=0.48\textwidth]{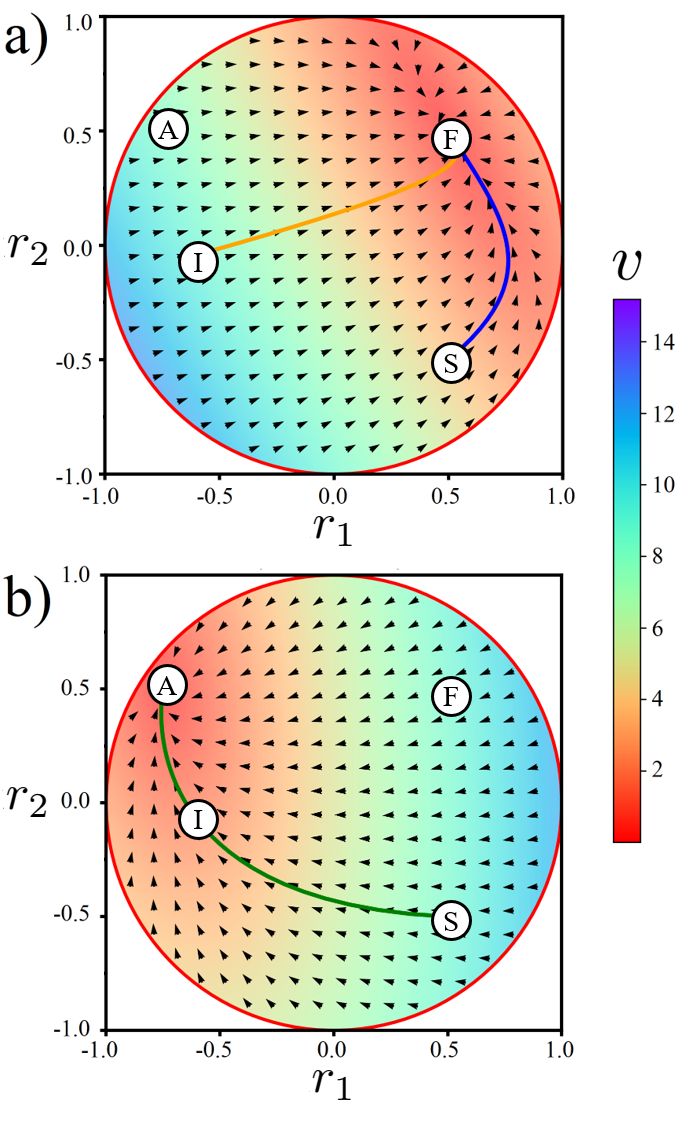}
\caption{ Velocity  field  $\dot{\rho}$ for the Markovian dynamics of a two-level system constrained to the $r_1$-$r_2$ plane ($r_3=0$) of the Bloch vector ${\bf r}(t)=(r_1,r_2,r_3)^T$ with $|{\bf r}|\le 1$. Here $\dot{\rho}$ can be represented by ${\bf v}({\bf r})=\dot{\bf r}$,  see Eq.~\eqref{v_vec}. Arrows and colors represent the direction and amplitude $(v=|{\bf v}|)$ of ${\bf v}({\bf r})$, respectively. In panel (a) [panel (b)], the steady state reached at long times is \textbf{F} [\textbf{A}], with the corresponding parameters in Eq.~\eqref{lindblad_first}  specified in the End Matter. 
The PME protocol compares the direct process ${\bf S}\to {\bf F}$ along trajectory $\Gamma_{\rm SF}$ [blue curve in panel (a)] to the two-step process composed of (i) $\bf{S}\to {\bf I}$ along trajectory $\Gamma_{\rm SA}$ [green curve in panel (b)] and (ii) ${\bf I} \to {\bf F}$ along $\Gamma_{\rm IF}$ [yellow curve in panel (a)].  If the two-step process is faster, the PME occurs. }    \label{fig1}
\end{figure}

Formally, we define the PME for arbitrary (classical or quantum, closed or open) systems as follows. We consider two system copies prepared in the \emph{same} initial state denoted by ${\bf S}$. At time $t=0$, the first copy is subjected to a control parameter quench driving the system toward the target state ${\bf F}$ in a time span $t_{\rm{SF}}$. 
The second copy is first subjected to a different parameter quench, corresponding to an auxiliary environment  (or Hamiltonian) driving the system toward the stationary state ${\bf A}$ within a time $t_{\rm{SA}}$. During the evolution toward ${\bf A}$, at a time $t_{\rm{SI}}<t_{\rm SA}$ where the intermediate state ${\bf I}$ has been reached, the system is decoupled from the auxiliary environment. By a second parameter quench, the system is then connected 
to the same environment used for the first system copy. The system will then reach the target state ${\bf F}$ in a time $t_{\rm{IF}}$.   The PME occurs if the condition $t_{\rm{SI}}+t_{\rm{IF}}<t_{\rm{SF}}$
is satisfied.  We illustrate this two-step PME protocol in Fig.~\ref{fig1}.
In the above example by Aristotle, ${\bf S}$ corresponds to the cold water,  
${\bf F}$ to ice, ${\bf A}$ is induced by the sun (or by fire) and would eventually lead to evaporated water, and ${\bf I}$ is the warm water.

While for the standard single-step ME, the initial state ${\bf I}$ is a stationary (and usually thermal) state, for the PME, there is no such constraint. Indeed, ${\bf I}$ is just  an arbitrary nonequilibrium state along the time evolution from ${\bf S}$ to ${\bf A}$ which offers  advantages in optimizing the relaxation speed. By employing methods from optimal control theory, one could systematically minimize the relaxation time by optimizing the choice of the states {\bf A} and ${\bf I}$.  Crucially, since both copies start from the same state ${\bf S}$, no parameter distance concept is needed anymore as in the standard single-step ME.  This is a key simplification which helps to avoid misinterpretations.

In order to quantify the PME, one needs a proper monitoring function in order to extract the relaxation times and thereby distinguish PME classes as for the single-step ME \cite{Lu2017,Nava2024}.  The monitoring function measures the distance of the time-dependent state from the target state and has to be a monotonically non-increasing, continuous,  and convex function of time \cite{Lu2017}. 
While our definition of the PME protocol and the classification reported below are generally valid, we illustrate the theory for Markovian quantum systems \cite{breuer2007theory} where the trace distance \cite{Nielsen2000} is a good monitoring function satisfying the above criteria.  The trace  distance has also been used in recent experiments demonstrating the ME \cite{Shapira2024,Zhang2025}.
However, other choices are possible, e.g., the Bures distance, and lead to the same PME classification, see also Ref.~\cite{Nava2024}, where the single-step ME was studied for Markovian quantum systems.  For closed quantum systems, one could
instead use the entanglement asymmetry as distance function~\cite{Ares2025}. 
Since we allow for multiple reservoirs coupled to the system, the stationary states can be current-carrying nonequilibrium steady states (NESSs). 

Another key concept introduced here is the \emph{velocity field}, which allows one to visualize in an intuitive manner if and how PMEs occur.  
The velocity field follows from the time derivative of the monitoring function. One can then accelerate relaxation processes by avoiding state space regions with small velocity field amplitude by designing suitable multi-step protocols.  
While the precise definition of the velocity field depends on the choice of the metric, the classification of different PME behaviors is independent of the metric (as long as it satisfies the above criteria).  Our analysis shows that all possible PMEs fall into three separate classes.  The theory is illustrated in detail for two-state Markovian quantum systems, where we demonstrate the emergence of all three classes.  We note that for large system dimension, the velocity field should be considered within a suitable hyperplane.

The above two-step protocol can be directly generalized to multi-step protocols, resulting in time-dependent Hamiltonian and Lindblad operators. This offers a promising perspective for optimal control theory \cite{Caneva2009,Koch_2016,Ansel_2024} in open systems, where also the Lindblad operators can be chosen, e.g., to optimize the relaxation speed.

\emph{Markovian dynamics.---}The first standard form of the Lindblad equation, which is the most general generator of Markovian dynamics, is ($\hbar=1$)
\begin{equation} \label{lindblad_first}
\frac{d\rho}{dt} = -i[H, \rho] + \sum_{m,n=1}^{N^2 - 1} C_{mn} \left( F^{}_m \rho F_n^\dagger - \frac{1}{2} \left\{ F_n^\dagger F^{}_m, \rho \right\} \right),
\end{equation} 
where $N$ is the system Hilbert space dimension, $\rho(t)$ the time-dependent density matrix, $H$  the system Hamiltonian, $\{F_m\}$ a basis of traceless operators in Hilbert space, 
$\{\cdot,\cdot\}$ the anticommutator, and $C$ the positive semi-definite Hermitian Kossakowski matrix \cite{Lindblad1976,breuer2007theory,Hayden_2022}.
By diagonalizing $C$, one can write Eq.~\eqref{lindblad_first} in the widely known diagonal form,
\begin{equation} \label{lindblad_diag}
\frac{d\rho}{dt} = -i[H, \rho] + \sum_{j=1}^{N^2 - 1} \gamma_j \left( L_j^{} \rho L_j^\dagger - \frac{1}{2} \left\{ L_j^\dagger L_j^{}, \rho \right\} \right),
\end{equation} 
where the jump operators $L_j$ are appropriate linear combinations of the operators $F_m$ and the rates $\gamma_j \geq 0$ are eigenvalues of $C$. The classical limit follows for $H\to 0$.
The diagonal form \eqref{lindblad_diag} uses the minimal number of jump operators; if the sum contains more than $N^2-1$ terms, it is in first standard form under disguise.  Since $C$ in Eq.~\eqref{lindblad_first} has dimension $(N^2-1)\times(N^2-1)$, there are up to $(N^2-1)^2$ independent dissipator parameters. Indeed, when resorting to Eq.~\eqref{lindblad_diag}, the freedom lies not just in the eigenvalues but also in the basis diagonalizing $C$.  
Similarly, the state can be expressed by $N^2-1$ parameters constrained by the positivity of $\rho$. We refer to the convex space spanned by all possible $\rho$ as the $\rho$-hyperplane. For a two-level system ($N=2$), it corresponds to the Bloch unit ball.  We employ the trace distance \cite{Nielsen2000},
$\mathcal{D}_T (\rho_1,\rho_2)=\frac12\mathrm{Tr}\left|\rho_1-\rho_2\right|$,
to measure the distance between density matrices $\rho_1$ and $\rho_2$ in the $\rho$-hyperplane. 

\emph{Velocity field.---}The classical thermal ME can be linked to geometric properties of a free energy landscape \cite{Lu2017}. However, this 
connection is not useful here since for a solution of Eqs.~\eqref{lindblad_first} or \eqref{lindblad_diag}, 
no energy landscape needs to be computed and NESS solutions generally do not realize minima of a free energy. 
We instead introduce a \emph{velocity field}\ landscape, which 
allows for an intuitive interpretation of the PME and directly depends on the jump operators and rates in Eq.~\eqref{lindblad_diag}. For Markovian systems, the relaxation dynamics in the $\rho$-hyperplane is represented in terms of
a velocity field by attaching to each $\rho(t)$ the vector $\dot{\rho}(t)$ (one may vectorize $\rho$ and $\dot \rho$ by using the Choi isomorphism \cite{Albert2014}.)
For given $H$ and $\{L_j\}$ in Eq.~\eqref{lindblad_diag},  $\dot{\rho}(t)$ is a function of $\rho(t)$ only. In Fig.~\ref{fig1},
we show the velocity field profile for a two-state system, with parameters such that all Bloch vectors ${\bf r}(t)$ have $r_3=0$. For $N>2$, one can visualize the velocity field in a hyperplane containing the points ${\bf S}, {\bf F},$ and ${\bf A}$ (or {\bf I}).  
For the two panels in Fig.~\ref{fig1}, we choose the Kossakowski matrices such that the corresponding steady states ${\bf F}$ and ${\bf A}$ with $\dot{\rho}=0$ are realized \cite{Diehl2008,Kraus2009}.  Both cases correspond to different environments and thus to different velocity field profiles.
In Fig.~\ref{fig1}(a), starting from the initial state \textbf{S}, the system evolves along a trajectory $\Gamma_{\rm SF}$ to the fixed point \textbf{F}, representing the first system copy in our PME definition. At each point, the velocity field is tangential to the trajectory. 
Comparing the trajectories with starting points \textbf{S} and \textbf{I} in Fig.~\ref{fig1}(a), we encounter a slower time evolution along  $\Gamma_{\rm SF}$ 
as compared to $\Gamma_{\rm IF}$ since a low-velocity region is traversed.  This
observation implies the standard ME.  Both time evolutions can be monitored by  $\mathcal{D}_T (\rho(t),\rho_{\rm F})$,   with $\rho(t)$ solving the Lindblad equation with $\rho(0)=\rho_{\rm S}$ and $\rho(0)=\rho_{\rm I}$, respectively. The trace distance will then smoothly decay to zero.  Since  \textbf{F}
is reached only for $t\to \infty$, we impose a finite but small dimensionless cutoff $\epsilon>0$ such that for $\mathcal{D}_{T}(\rho(t_c),\rho_{\rm F})=\epsilon$, convergence is declared for $t>t_c$. For $\epsilon\ll 1$, the PME classification below is independent of the precise value of $\epsilon.$

The velocity field $\dot{\rho}$ offers an intuitive understanding of the PME. Following  our PME definition,  we compare the velocity fields for two different environments, namely the one corresponding to the target state \textbf{F}, see Fig.~\ref{fig1}(a), and the one for the auxiliary state \textbf{A}, see Fig.~\ref{fig1}(b).  For both system copies, we start from the initial state \textbf{S}. 
The first copy evolves in the velocity field of the target state only, see the blue curve in Fig.~\ref{fig1}(a), 
and thus has to propagate through a low-velocity (``slow'') region. 
The second copy instead first evolves in the velocity field of the auxiliary environment toward \textbf{A},
see the green  curve in Fig.~\ref{fig1}(b), where the propagation speed is much faster and the slow region in Fig.~\ref{fig1}(a)
is efficiently circumvented.  To approach the target state \textbf{F}, at some intermediate state \textbf{I} along the 
trajectory $\Gamma_{\rm SA}$, reached at time $t_{\rm SI}<t_{\rm SA}$, the system is decoupled from the auxiliary environment and connected to the target environment by  another parameter quench. The system then evolves along the trajectory $\Gamma_{\rm IF}$, see the yellow curve in Fig.~\ref{fig1}(a). 
In fact, by avoiding the slow region in Fig.~\ref{fig1}(a), it reaches \textbf{F} before the first copy and  the PME occurs. 

\begin{figure}
\includegraphics[width=0.4\textwidth]{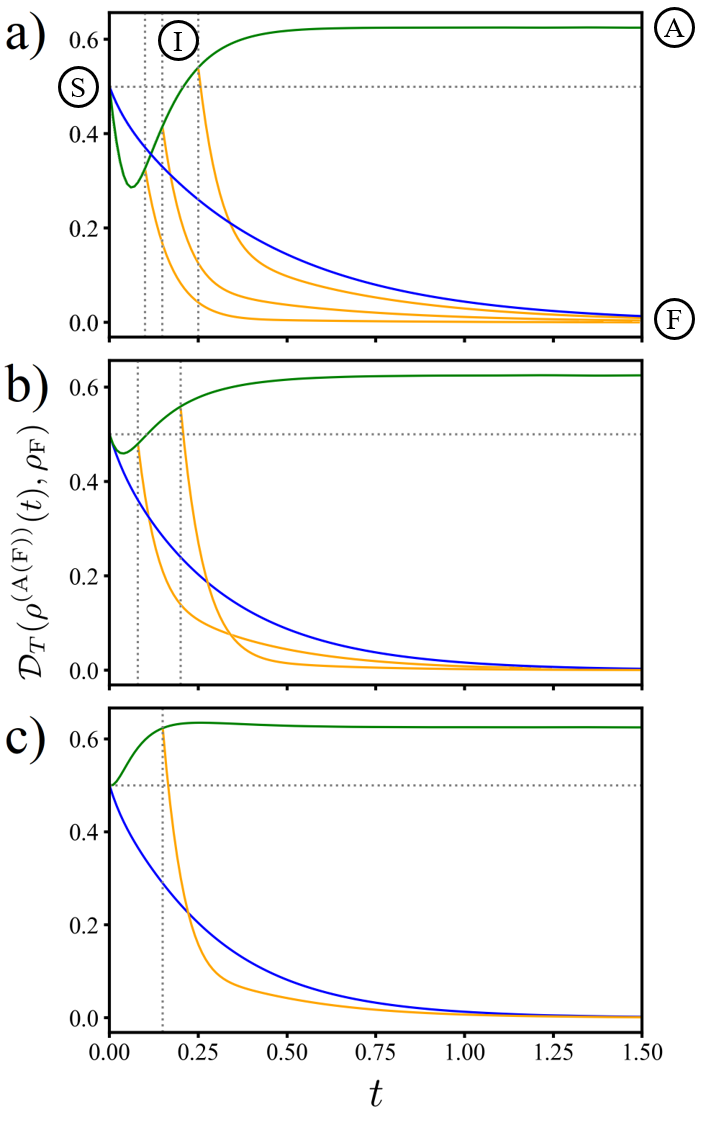}
\caption{Trace distance ${\cal D}_T (\rho^{({\rm F}\, ({\rm A}))}(t), \rho_{\rm F})$  vs time $t$ using different target (\textbf{F})  and auxiliary (\textbf{A}) states as environment for Markovian two-state systems. As in Fig.~\ref{fig1}, blue (green) curves show the time evolution of ${\cal D}_T (\rho^{({\rm F} \,({\rm A}))}(t), \rho_{\rm F})$ along  ${\bf S}\to {\bf F}({\bf A})$ under the influence of the respective environment, and yellow curves show ${\cal D}_T (\rho^{({\rm F})}(t), \rho_{\rm F})$ along ${\bf I}\to {\bf F}$ for
different intermediate states \textbf{I}.  Dotted vertical and horizontal lines are guides to the eyes only. For parameter values, see End Matter.
{\bf (a):} Example where ${\cal D}_T (\rho^{({\rm A})}(t), \rho_{\rm F})$ has a minimum and a crossing point with ${\cal D}_T (\rho^{({\rm F})}(t), \rho_{\rm F})$. 
For three states \textbf{I}, yellow curves show the trace distance along $\Gamma_{\rm IF}$.  These three cases realize all three PME types. {\bf (b):} Example where ${\cal D}_T (\rho^{({\rm A})}(t), \rho_{\rm F})$ has a minimum but no crossing point with ${\cal D}_T (\rho^{({\rm F})}(t), \rho_{\rm F})$.  
{\bf (c):} Example for the case in Eq.~\eqref{case-3}, where no minimum in ${\cal D}_T (\rho^{({\rm A})}(t), \rho_{\rm F})$ exists. 
}  \label{fig2}
\end{figure}

\emph{PME classification.---}We first analyze the relative position between the trajectories $\Gamma_{\rm SF}$ and $\Gamma_{\rm SA}$. A first case arises if, starting from ${\bf S}$, the system dynamics toward {\bf A} (indicated by the superscript ${}^{({\rm A})}$ below) 
initially reduces the trace distance to both states \textbf{A} and \textbf{F}, while at later times the trajectory moves away from \textbf{F}. 
This first case is therefore characterized by the two conditions
\begin{eqnarray} \label{case-1}
& &\partial_t \mathcal{D}_T (\rho^{({\rm A})}(t=0),\rho_{\rm F})<0 , \nonumber \\
& &\exists \: t_M>0 \: \mid \: \partial_t \mathcal{D}_T (\rho^{({\rm A})}(t=t_M+0^+),\rho_{\rm F})>0.
\end{eqnarray}
The second condition means that $\mathcal{D}_T (\rho^{({\rm  A})}(t),\rho_{\rm F})$ has a minimum at  $t=t_M$ corresponding to the state along $\Gamma_{\rm SA}$ closest to \textbf{F}.  If there are several minima, $t_M$ refers to the first minimum along $\Gamma_{\rm SA}$.
We further distinguish two sub-cases according to the conditions 
\begin{equation}\label{case-1-subcases}
\partial_t \mathcal{D}_T (\rho^{({\rm A})}(t=0),\rho_{\rm F})\lessgtr 
\partial_t \mathcal{D}_T (\rho^{({\rm F})}(t=0),\rho_{\rm F}),
\end{equation}
illustrated in Figs.~\ref{fig2}(a) and \ref{fig2}(b), respectively.
These two sub-cases are characterized by the existence, or not, of a finite-time crossing point between the trace distance curves 
$\mathcal{D}_T (\rho^{({\rm A})}(t),\rho_{\rm F})$ and $\mathcal{D}_T (\rho^{({\rm F})}(t),\rho_{\rm F})$, 
where the superscript ${}^{({\rm F})}$ means that the dynamics is driven toward the target state ${\bf F}$.
A second case arises if $\mathcal{D}_T (\rho^{({\rm A})}(t),\rho_{\rm F})$ monotonically decreases in time. It is similar to the first case but the minimum along $\Gamma_{\rm SA}$ in the trace distance to \textbf{F} is now at $t_M \to \infty$; we do not discuss this case further. 
Finally, a third case exists if, starting from \textbf{S}, the system moves away from \textbf{F}, thus increasing the trace distance from it, while being attracted to \textbf{A}. One then arrives at the condition
\begin{equation} \label{case-3}
\partial_t \mathcal{D}_T (\rho^{({\rm A})}(t=0),\rho_{\rm F})>0.
\end{equation}
An example is given in Fig.~\ref{fig2}(c). 

We now provide the general PME classification. For the second system copy, the intermediate state \textbf{I} is reached at  time $t_{\rm SI}$ along the trajectory $\Gamma_{\rm SA}$.  
One now performs a parameter quench to decouple the system from the auxiliary (${\bf A}$) environment and instead couples it to the target (${\bf F}$) environment, i.e., 
$\rho(t<t_{\rm SI})=\rho^{({\rm A})}(t)$ and $\rho(t\geq t_{\rm SI})=\rho^{({\rm F})}(t)$.
We then distinguish three different PME types depending on the location of the corresponding intermediate state \textbf{I}.  If the condition
\begin{equation} \label{weak-1}
\mathcal{D}_T (\rho^{({\rm A})}(t_{\rm SI}),\rho_{\rm F})<
\mathcal{D}_T (\rho^{({\rm F})}(t_{\rm SI}),\rho_{\rm F})
\end{equation}
holds, we define the \textit{weak type-A PME}.  An example is given by the 
lowest-lying yellow curve in Fig.~\ref{fig2}(a). It is characterized by the absence of a crossing point between the trace distances for the first and the
second system copy, with the state of the second copy being closer to ${\bf F}$ at all times. 
On the other hand, for
\begin{equation} \label{weak-2}
\mathcal{D}_T (\rho_{\rm S},\rho_{\rm F})>\mathcal{D}_T (\rho^{({\rm A})}(t_{\rm SI}),\rho_{\rm F})>\mathcal{D}_T (\rho^{({\rm F)}}(t_{\rm SI}),\rho_{\rm F}),
\end{equation}
we speak of a \textit{weak type-B PME}. Examples are given by the middle yellow curve in Fig.~\ref{fig2}(a) and by the lower yellow curve in Fig.~\ref{fig2}(b), with a finite-time crossing point between both trace distances.  
Finally, the most elusive scenario, dubbed \textit{strong PME}, may arise for the third case above, see Eq.~\eqref{case-3}, if
$\mathcal{D}_T (\rho^{({\rm A})}(t_{\rm SI}),\rho_{\rm F})>\mathcal{D}_T (\rho_{\rm S},\rho_{\rm F}).$
Examples for the \textit{strong} PME are given by the upper yellow curves in Figs.~\ref{fig2}(a,b) and by the yellow curve in Fig.~\ref{fig2}(c). This is the hardest and most counterintuitive PME to achieve, but it is closest in spirit to the Pontus example by Aristotle \cite{Aristotle}. From an application point of view, however, we expect that the two \textit{weak} PME cases will provide faster relaxation times toward \textbf{F}. 

\emph{Two-level systems.---}As probably simplest application, we now turn to Markovian two-level systems.  The results in Figs.~\ref{fig1} and \ref{fig2} were obtained for this case and show that all three PME classes can already be realized for $N=2$.  For details and the parameters in Figs.~\ref{fig1} and \ref{fig2}, see End Matter.
For $N=2$, we express the traceless operators $F_n$ in Eq.~\eqref{lindblad_first} by the Pauli matrices $\sigma_{n=1,2,3}$, and the state is $\rho(t)=\frac{1}{2}(\mathbb{I}+\sum_n r_{n}(t) \sigma_{n})$ with the Bloch vector ${\bf r}(t)$. The trace distance between  $\rho_1$ and $\rho_2$ with Bloch
vectors ${\bf r}_1$ and ${\bf r}_2$ is then given by ${\cal D}_T(\rho_1,\rho_2)=\frac12| {\bf r}_1-{\bf r}_2|$.  Furthermore, in Eq.~\eqref{lindblad_first},
the Hamiltonian is $H=\sum_n h_n \sigma_n$ with a 
real-valued vector ${\bf h}=(h_1,h_2,h_3)^T$ and the Kossakowski matrix 
$
C=\left(\begin{array}{ccc}
C_{11} & C_{12} & C_{31}^{*}\\
C_{12}^{*} & C_{22} & C_{23}\\
C_{31} & C_{23}^{*} & C_{33}
\end{array}\right),
$ 
with real-valued diagonal elements and three independent
complex-valued off-diagonal entries.
The velocity field  
$\dot \rho$ is written in Bloch representation as vector field ${\bf v}({\bf r})$, where the Lindblad equation gives
\begin{equation} \label{v_vec}
\dot{\bf r} = {\bf v}({\bf r})=2 {\bf \Lambda} \cdot {\bf r} + {\bf b} .
\end{equation}
The $3\times 3$ matrix ${\bf \Lambda}$ and the vector ${\bf b}$ are expressed in terms of $C$ and ${\bf h}$.  
Using these equations, we obtain Figs.~\ref{fig1} and \ref{fig2} and the corresponding PME types.
For $N>2$, similar representations with ${\bf v}=\dot{\bf r}$ 
apply by using the $(N^2-1)$-dimensional coherence vector ${\bf r}$ generalizing the Bloch vector \cite{Kasatkin2023,Tscherbul2024}.
All three PME classes could thus be realized experimentally in Markovian two-state systems.  There are various physical platforms to achieve this goal with reservoir engineering techniques, 
e.g., trapped ions \cite{Joshi2024,Shapira2024}, superconducting qubits \cite{Rasmussen2021}, or ultracold atoms \cite{Bloch2008}.
Using the $C$ matrices in the End Matter, by diagonalizing $C=U^{\dagger}\Lambda U$, where the unitary $U$ contains the eigenvectors
and the diagonal matrix $\Lambda$ the eigenvalues, the jump operators follow as $L_j=\sum_i U_{ji} \sigma_i$ with rates $\gamma_j=\Lambda_{jj}$ in Eq.~\eqref{lindblad_diag}.  The time-dependent trace distance can be measured by quantum state tomography.
We conclude that an experimental realization of the PME and a test of our 
classification into three separate PME types seems in direct reach.

\emph{Conclusions.---}We introduced the PME as a generalized Mpemba protocol taking into account the time needed to prepare the
initially ``far'' state. It is conceptually simpler and advantageous for applications compared to the single-step Mpemba protocol.    We find three PME classes which have been illustrated explicitly for  Markovian two-state quantum systems.  We note that
non-Markovian effects \cite{Breuer2016,Strachan2024} can help to further speed up the PME relaxation dynamics, see also Ref.~\cite{Deffner2013}. Appropriate distance functions could be constructed by resorting to quantum resource theories \cite{Chitambar2019}.  We leave this point to future work, as well as the systematic optimization of the states ${\bf A}$ and ${\bf I}$ by means of optimal control theory \cite{Caneva2009,Koch_2016,Ansel_2024} and the generalization to multi-step protocols, e.g., for a study of optimal heat engines. Let us also note that for large system space dimensions,  
additional simplifications may be necessary to reduce the computational demands.

\begin{acknowledgments} 
We thank D. Giuliano and X. Turkeshi for discussions and  acknowledge funding by the Deutsche Forschungsgemeinschaft (DFG, German Research Foundation) under Projektnummer 277101999 - TRR 183 (project B02) and under Germany's Excellence Strategy - Cluster of Excellence Matter and Light for Quantum Computing (ML4Q) EXC 2004/1 - 390534769.
\end{acknowledgments}

\emph{Data availability.---}Numerical data used for generating the figures are available at the Zenodo site
\new{https://doi.org/10.5281/zenodo.17143780}

\bibliography{biblio}

\appendix
\section*{End Matter}
\setcounter{equation}{0}
\setcounter{figure}{0}
\renewcommand{\theequation}{A\arabic{equation}}
\renewcommand{\thefigure}{A\arabic{figure}}

We here provide technical details concerning the two-state realization of the PME.  First, we note that one can perform a row-wise vectorization of the Lindblad equation by means of the Choi isomorphism \cite{Albert2014}, using the relation
$\left|\left.ABC\right\rangle \right\rangle =\left(A\otimes C^{T}\right)\left|\left.B\right\rangle \right\rangle$,
where $\otimes$ is the Kronecker product and $\left|\left.B\right\rangle \right\rangle$ the vectorized form of an $N\times N$ matrix $B$ obtained by stacking matrix rows into an $N^2$-dimensional column vector. 
For $N=2$, the vectorized form of Eq.~\eqref{lindblad_first} is $\left|\left.\dot{\rho}\right\rangle \right\rangle =\mathcal{L}\left|\left.\rho\right\rangle \right\rangle$ with 
\begin{align}
\mathcal{L} & =-i\left(H\otimes\mathbb{I}-\mathbb{I}\otimes H^{T}\right)+\sum_{m.n=1}^{3}C_{mn}\Bigl[\left(\sigma_{m}\otimes\sigma_{n}^{T}\right)\nonumber \\
 &-\frac{1}{2}\left(\sigma_{n}\sigma_{m}\otimes\mathbb{I}\right)-\frac{1}{2}\left(\mathbb{I}\otimes\sigma_{m}^{T}\sigma_{n}^{T}\right)\Bigr].
\end{align}
In terms of the Bloch vector ${\bf r}$, we obtain the dynamical equation \eqref{v_vec}
with the matrix
\begin{equation} \label{Lambda_matrix}
{\bf \Lambda}=\left(\begin{array}{ccc}
-C_{22}-C_{33} & -h_{3}+\mathrm{Re}C_{12} & h_{2}+\mathrm{Re}C_{31}\\
h_{3}+\mathrm{Re}C_{12} & -C_{11}-C_{33} & -h_{1}+\mathrm{Re}C_{23}\\
-h_{2}+\mathrm{Re} C_{31} & h_{1}+\mathrm{Re}C_{23} & -C_{11}-C_{22}
\end{array}\right)
\end{equation}
and the vector
\begin{equation} \label{b_vector}
{\bf b}=-4 \, {\rm Im}\, \left(C_{23}, C_{31},C_{12}\right)^T.
\end{equation}
These quantities depend on the Kossakowski matrix $C$ and on the 
vector ${\bf h}$ defining the system Hamiltonian.
Let us briefly show that the trace distance between two density matrices $\rho_1$ and $\rho_2$ is proportional to the Euclidean distance of the corresponding Bloch vectors ${\bf r}_1$ 
and ${\bf r}_2$. Indeed,
$\mathcal{D}_T(\rho_1,\rho_2)=\frac{1}{2}\sum_{i}\left|\lambda_{i}\right|$,
where $\lambda_i$ are the eigenvalues of 
$\rho_{1}-\rho_{2}=\frac{1}{2}\left(\begin{array}{cc}
\Delta r_{3} & \Delta r_{1}-i\Delta r_{2}\\
\Delta r_{1}+i\Delta r_{2} & -\Delta r_{3}
\end{array}\right)$ with $\Delta r_m\equiv r_{1,m}-r_{2,m}$.  One finds
$\lambda_i=\pm\frac{1}{2}\sqrt{\Delta r_{1}^{2}+\Delta r_{2}^{2}+\Delta r_{3}^{2}}$,
which directly implies 
$\mathcal{D}_T(\rho_1,\rho_2)=\frac{1}{2}|{\bf r}_1-{\bf r}_2|$.
From the trace distance, we extract the scalar velocity,
\begin{equation}
    v(t) = 2\lim_{\delta t \to 0} \frac{\mathcal{D}_T \left( \rho(t + \delta t), \rho(t) \right)}{\delta t} = \mathrm{Tr} \left| \dot{\rho}(t) \right| =  |\dot{\bf r}(t)|.
\end{equation}

Going back to Eq.~\eqref{v_vec}, we next consider a dissipative two-state dynamics confined to the $r_1$-$r_2$ plane of the Bloch ball, see Fig.~\ref{fig1}, by judiciously choosing parameters enforcing $r_3(t)= 0$. To that end, we set
\begin{equation}
\mathrm{Im}C_{12}= -h_2+\mathrm{Re}C_{31}=h_1+\mathrm{Re}C_{23}=0
\label{conditionr30}
\end{equation}
in Eqs.~\eqref{Lambda_matrix} and \eqref{b_vector}. For $r_3(0)=0$, one can then write 
Eq.~\eqref{v_vec} as
$
\partial_{t}\left(\begin{array}{c}
r_{1}\\r_{2}
\end{array}\right)=2{\bf \Lambda}'\cdot \left(\begin{array}{c}
r_{1}\\ r_{2}
\end{array}\right)+{\bf b}'   
$
with 
${\bf \Lambda}'=\left(\begin{array}{cc}
-C_{22}-C_{33} & -h_{3}+\mathrm{Re}C_{12}\\
h_{3}+\mathrm{Re}C_{12} & -C_{11}-C_{33}
\end{array}\right)$
and  
${\bf b}'=-4 \, {\rm Im}\, \left(C_{23},C_{31}\right)^T$.
The entries in ${\bf \Lambda}'$ and ${\bf b}'$ are constrained by the condition 
that $C$ is a physically allowed (Hermitian and positive semi-definite) Kossakowski matrix. 
In practice, it is more convenient to choose the stationary state ${\bf r}_*$ 
rather than the vector ${\bf b}'$, using the relation ${\bf b}'=-2{\bf \Lambda}'\cdot {\bf r}_*$.

\begin{figure}
\includegraphics[width=0.4\textwidth]{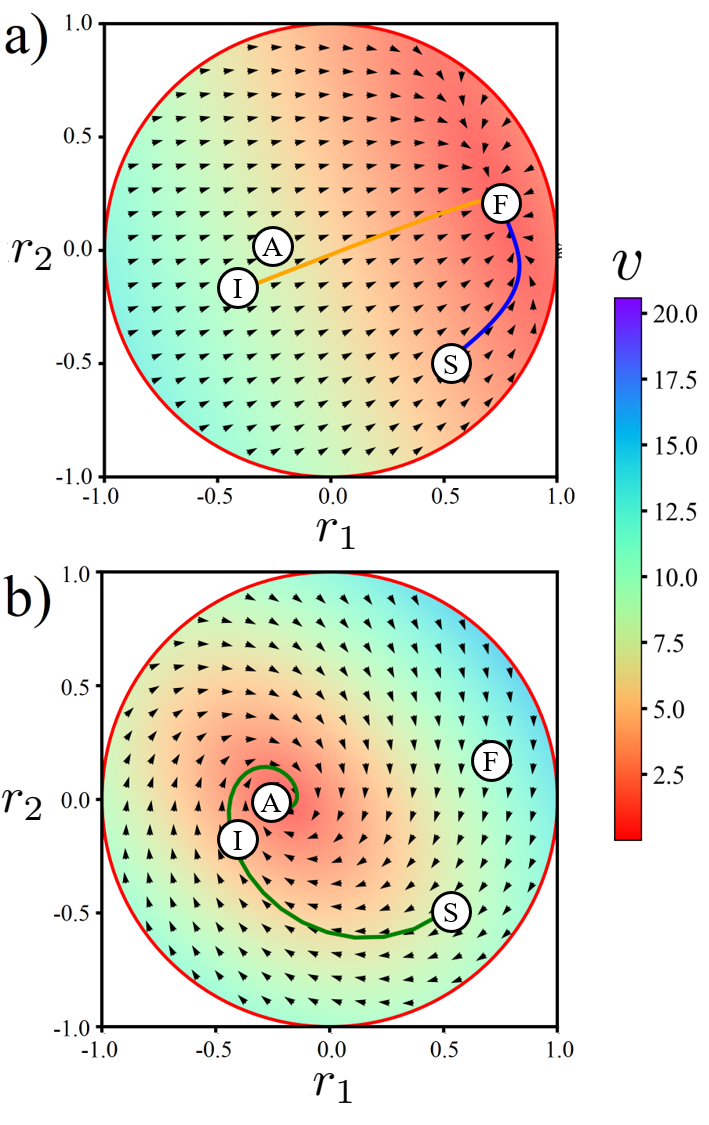}
\caption{Velocity  field profiles for a two-level system constrained to the $r_1$-$r_2$ plane as in Fig.~\ref{fig1}, with the Kossakowski matrices \eqref{quantum_kossakowski}. {\bf (a):} Classical dissipative dynamics induced by the target bath without Hamiltonian contribution, ${\bf h}^{({\rm F})}=0$. {\bf (b):} Strong quantum effects due to a large Hamiltonian contribution, ${\bf h}^{({\rm A})}=(0,0,-10)^T$, resulting in a spiral-type trajectory $\Gamma_{\rm SA}$.
}  \label{figA1}
\end{figure}

\begin{figure}
\includegraphics[width=0.48\textwidth]{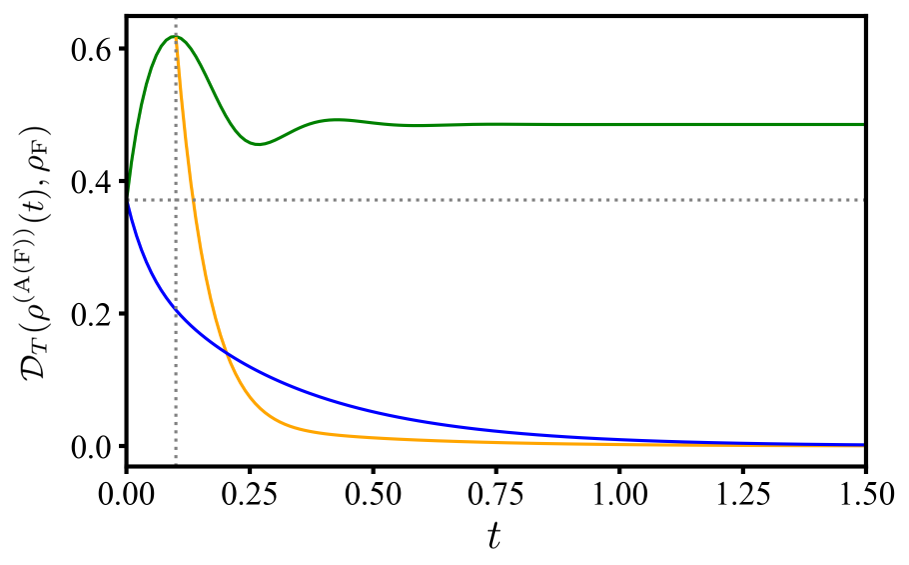}
\caption{Trace distances ${\cal D}_T (\rho^{({\rm F}\, ({\rm A}))}(t), \rho_{\rm F})$  
vs time $t$ as in Fig.~\ref{fig2} but for the parameters in Fig.~\ref{figA1}.
The blue (green) curves show the time evolution of ${\cal D}_T (\rho^{({\rm F}\, ({\rm A}))}(t), \rho_{\rm F})$ along  ${\bf S}\to {\bf F} ({\bf A})$ under the influence 
of the respective environment, and the yellow curve shows ${\cal D}_T (\rho^{({\rm F})}(t), \rho_{\rm F})$ along ${\bf I}\to {\bf F}$ for
an intermediate state \textbf{I}.  Dotted vertical and horizontal lines are guides to the eyes only.
}
 \label{figA2}
\end{figure}

We then summarize the parameters used in Figs.~\ref{fig1} and \ref{fig2}.  With $r_3=0$, we set ${\bf r}_{\rm S}=(0.5,-0.5)^T$ for the initial state \textbf{S}, ${\bf r}_{\rm F}=(0.5,0.5)^T$ for the target state \textbf{F}, and ${\bf r}_{\rm A}=(-0.75,0.5)^T$ for the auxiliary state ${\bf A}$. In Fig.~\ref{fig2}(a), the Kossakowski matrices for the target and auxiliary environments were taken as
\begin{align} \nonumber
C^{({\rm F})}= & \left(\begin{array}{ccc}
1 & -2 & 0.94i\\
-2 & 5 & -2.06i\\
-0.94i & 2.06i & 1
\end{array}\right),\\
C^{({\rm A})}= & \left(\begin{array}{ccc}
4.5 & -2 & 1.75i\\
-2 & 3 & 1.06i\\
-1.75i & -1.06i & 2.5
\end{array}\right),
\end{align}
using ${\bf h}^{({\rm F})}=(0,0,0.25)^T$ and ${\bf h}^{({\rm A})}=(0,0,2)^T$ while $t_{\rm SI}= 0.1$ (\textit{weak type-A PME}), $t_{\rm SI}= 0.15$ (\textit{weak type-B PME}), and $t_{\rm SI}= 0.25$ (\textit{strong PME}).
For Fig.~\ref{fig2}(b), we instead took
\begin{align}
C^{({\rm F})}= & \left(\begin{array}{ccc}
1 & -2 & 1.06i\\
-2 & 5 & -2.19i\\
-1.06i & 2.19i & 1.5
\end{array}\right),\nonumber\\
C^{({\rm A})}= & \left(\begin{array}{ccc}
2 & -1 & 0.59i\\
-1 & 5 & 2.12i\\
-0.59i & -2.12i & 1.5
\end{array}\right),
\end{align}
with ${\bf h}^{({\rm F})}={\bf h}^{({\rm A})}=(0,0,0.25)^T$ and $t_{\rm SI}= 0.08$ 
(\textit{weak type-B PME}), and $t_{\rm SI}= 0.2$ (\textit{strong PME}).
Finally, for Fig.~\ref{fig1} and Fig.~\ref{fig2}(c), we used
\begin{align}
C^{({\rm F})}= & \left(\begin{array}{ccc}
1 & -2 & 0.88i\\
-2 & 5 & -2.38i\\
-0.88i & 2.38i & 1.5
\end{array}\right),\nonumber\\
C^{({\rm A})}= & \left(\begin{array}{ccc}
2 & -1 & -0.06i\\
-1 & 5 & 2.56i\\
0.06i & -2.56i & 1.5
\end{array}\right),
\end{align}
with ${\bf h}^{({\rm F})}=(0,0,1)^T$ and ${\bf h}^{({\rm A})}=(0,0,-1.5)^T$ with $t_{\rm SI}= 0.15$ (\textit{strong PME}).
The above parameter choices are consistent with Eq.~\eqref{conditionr30}.

Next we provide a comparison between a purely dissipative dynamics
and a quantum-dominated dynamics due to a large Hamiltonian part in Eq.~\eqref{lindblad_first}. In Fig.~\ref{figA1}, we show the velocity field profiles for a Markovian two-state system as in Fig.~\ref{fig1}.
We here set ${\bf r}_{\rm S}=(0.5,-0.5)^T$, ${\bf r}_{\rm F}=(0.75,0.2)^T$, and 
${\bf r}_{\rm A}=(-0.2,0)^T$. The Kossakowski matrices were taken as
\begin{align} \label{quantum_kossakowski}
C^{({\rm F})}= & \left(\begin{array}{ccc}
1 & -2 & i\\
-2 & 5 & -2.64i\\
-i & 2.64i & 1.5
\end{array}\right),\nonumber\\
C^{({\rm A})}= & \left(\begin{array}{ccc}
5 & -1 & -1.1i\\
-1 & 1 & 0.25i\\
1.1i & -0.25i & 1.5
\end{array}\right).
\end{align}
In Fig.~\ref{figA1}(a), we set ${\bf h}^{({\rm F})}=0$ such that the dynamics toward the target state \textbf{F} is purely dissipative (classical). 
We observe that the trajectories $\Gamma_{\rm SF}$ and $\Gamma_{\rm IF}$ are curved contractive paths, which are shaped by the Kossakowski matrix $C^{({\rm F})}$ only. In Fig.~\ref{figA1}(b), we instead set ${\bf h}^{({\rm A})}=(0,0,-10)^T$ such that we have a large Hamiltonian contribution governing the dynamics toward the state \textbf{A}.  Indeed, the trajectory $\Gamma_{\rm SA}$ is a curve approaching \textbf{A} for $t\to \infty$ with damped spiral oscillations. We note that for $C^{({\rm A})}\to 0$, the resulting purely unitary dynamics for ${\bf r}(t)$ describes a circle around the origin. 

Finally, in Fig.~\ref{figA2}, we show the corresponding time-dependent trace distances $\mathcal{D}_T (\rho^{({\rm F \,(A)})}(t),\rho_{\rm F})$ for the direct process ${\bf S}\to {\bf F}$ along $\Gamma_{\rm SF}$ and for the two-step process composed of $\bf{S}\to {\bf I}$ along $\Gamma_{\rm SA}$ followed by 
${\bf I} \to {\bf F}$ along $\Gamma_{\rm IF}$ starting at $t_{\rm SI}= 0.1$. 
For a quantum-dominated dynamics, $\mathcal{D}_T (\rho^{({\rm A})}(t),\rho_{\rm F})$ exhibits damped oscillations before reaching the asymptotic $t\to \infty$ value determined by the state ${\bf A}$.  We emphasize that $\mathcal{D}_T (\rho^{({\rm F})}(t),\rho_{\rm F})$ and $\mathcal{D}_T (\rho^{({\rm A})}(t),\rho_{\rm A})$ are always 
monotonically non-increasing functions of time under Markovian evolution, 
even in the presence of a quantum Hamiltonian.

\end{document}